\documentstyle[preprint,aps]{revtex}

\begin{document}

\draft

\title{Master equation for a particle coupled to a two-level reservoir}

\author{Paulo C. Marques$^{1}$ \ and \ A.H. Castro Neto$^{2,3}$}
\address{$^{1}$ Instituto de F\'\i sica da Universidade de S\~ao Paulo \\
C.E.P. 01452-990, C.P. 20516, S\~ao Paulo, S.P., Brazil.}
\address{$^{2}$ Institute for Theoretical Physics\\
University of California in Santa Barbara\\
Santa Barbara, CA, 93106-4030}
\address{$^{3}$ Department of Physics\\
University of Illinois at Urbana-Champaing\\
1110 W. Green St, Urbana, IL, 61801}
\maketitle

\begin{abstract}

We study the quantum dissipative dynamics of a particle coupled
linearly to a set of two-level systems (the heat bath) via the master
equation method which we extract from the path integral formalism
independently from the form of the bath spectral density.
We compare our results with the standard models
based on bosonic heat baths showing their main differences and
similarities. In particular, we study special forms for the
spectral density of the bath which give results quite different from the
standard models.

\end{abstract}

\vspace{15mm}


The study of quantum open systems is one of the most important issues
of statistical quantum mechanics because of its application to many
different subjects such as thermalization (tendency to thermal equilibrium),
lost of quantum coherence and quantum brownian motion. This kind of
theoretical approach has been used with success in order to study
macroscopic quantum tunneling \cite{caldeira,spinboson}, evolution of
the early universe \cite{hartle} and measurement theory \cite{zurek}, among
others.

The field of quantum brownian motion had a major revival after the
work of Caldeira and Leggett \cite{caldeira} on the study of a particle
coupled to a set of harmonic oscillators via the method devised by
Feynman and Vernon \cite{feynman} using path integrals. The path integral
method allows the trace over the oscillators variables (the heat bath) and
the study of the coordinate of interest. As it is already well known
the dissipation and diffusion of a particle coupled linearly to a heat
bath depend essentially on the spectral density, $J(\omega)$, of the heat bath
(which is related to the response function of the reservoir at some
frequency $\omega$). In the Caldeira-Leggett
model \cite{caldeira} the dissipation is supposed to be ohmic, that is,
the spectral density of the bath has the form,
\begin{equation}
J_0(\omega) \ = \ \eta_0 \; \omega \ \theta (\omega_c - \omega),
\end{equation}
where $\omega_c$ is a suitable cut-off frequency. With this special choice
for the spectral density Caldeira and Leggett showed that the semiclassical
dynamics of the system is described by the Langevin equation with a
dissipation coefficient $\eta_0$.

However, the study of open quantum systems has a long history with many
branches. One of the main streams of research is
based on the study of master equations (such as the Fokker-Planck equation)
for the dynamical evolution of quantum systems which is used in
quantum optics \cite{louisell,zwanzig}. In the seventies, Lindblad
gave a general mathematical criteria for the classification and existence of
such master equations \cite{lindblad}. More recently, the methods of master
equations has been used extensively in the literature \cite{unruh,hu,gallis}.
Actually, Caldeira and Leggett obtained the equation
for the evolution of the reduced density matrix, $\rho_s$,
(the density matrix after the trace over the oscillators bath)
with the ohmic spectral density $(1)$ and in the high temperature limit
($k_B T \gg \hbar \omega_c$) which, in the position
representation ($\rho_s(x,y,t) = \langle x | \rho_s(t) | y \rangle$), reads
\cite{caldeira},
\begin{eqnarray}
i\hbar \frac{d{\rho_s}(x,y,t)}{dt} \ &=&
\ \Biggl[ \ -\frac{\hbar^2}{2m} \Biggl(\frac{\partial^2}{\partial x^2}
- \frac{\partial^2}{\partial y^2}\Biggl)
\ - \ i\hbar \; \Gamma \; (x - y)\left(\frac{\partial}{\partial x}
- \frac{\partial}{\partial y}\right) \nonumber \\
&+& \ V_R(x) \ - \ V_R(y)
\ - \ i \ D_{pp}\; (x - y)^2 \ \Biggr] \; {\rho_s}(x,y,t),
\end{eqnarray}
where $m$ is the mass of the particle, $V_R$ is the
renormalized external potential and
\[
D_{pp} \ = \ \frac{2 \eta_0 k_B T}{\hbar}\ ,
\ \ \ \ \ \ \ \ \ \
\Gamma \ = \ \frac{\eta_0}{m}\ ,
\]
are the diffusion coefficient in momentum space and the effective dissipation
coefficient, respectively.

In this paper we address the problem of the master equation for a
different kind of bath which was proposed recently in the literature
by Caldeira {\it et al.} \cite{twolevel} and has distinct features
from the models described above. In the model proposed in Ref.\cite{twolevel}
the particle interacts with a heat bath composed of two-level systems.
The hamiltonian of interest can be written as a sum of three terms,
\[
H \ = \ H_s \ + \ H_i \ + \ H_b
\]
where $H_s$ is the Hamiltonian for a particle in an external potential
$V(x)$,
\[
H_s\ =\ \frac{p^2}{2 m} \ + \ V(x),
\]
$H_i$ gives the interaction between the particle and the bath,
\[
H_i \ = \ - \ \sum_n \ J_n \; x \; \sigma_{x_n},
\]
and the bath hamiltonian is,
\[
H_b \ = \ \sum_n \ \frac{\hbar \omega_n}{2} \; \sigma_{z_n}
\]
where $\sigma_{x_n}$ and $\sigma_{z_n}$ are the usual Pauli matrices and
$J_n$ is the coupling constant for the particle with the ${\rm n^{th}}$
two-level system.

As it was shown in \cite{twolevel}, the dynamics of the
particle can be evaluated in the path integral formalism via the
superpropagator, $J$, \cite{feynman}. Assuming that initially at some time
$t=0$ the particle and the bath are decoupled the time evolution of the
system can be written as,
\begin{equation}
\rho_s (x, y, t)
= \int_{-\infty}^{+\infty}{dx_0}
\int_{-\infty}^{+\infty}{dy_0}
\; J(x, y, t \ | \ x_0, y_0,0) \; \rho_s(x_0,y_0,0)
\end{equation}
where the superpropagator is given by,
\begin{equation}
J(t,0) \ = \ \int^{x}_{x_0}{{\cal D}x(s)}
\int^{y}_{y_0}{{\cal D}y(s)} \, \,
{\rm exp}\ \frac{i}{\hbar}\ \biggl[ S\left[ x(s) \right]
- S\left[ y(s) \right] \biggl] \; F\left[ x(s) , y(s) \right].
\end{equation}
In the last equation, $F\left[ x(s) ,y(s) \right]$ is the influence
functional \cite{feynman}, and
$S\left[ x(s) \right]$ is the free action for the particle,
\[
S\left[ x(s) \right]
\ = \ \int^t_0{ds} \left[ \; \frac{m \dot{x}^2(s)}{2}
\; - \; V\left[ x(s) \right]\; \right]\ .
\]
In this paper we will be interested in the case where the external potential
is harmonic, that is,
$V\left[ x(s) \right] \; = \; \frac{1}{2}\, m\, \Omega_0^2\, x^2(s) $.

In the weak coupling limit the influence functional reads \cite{twolevel},
\begin{eqnarray}
F[q(s),Q(s)] \ &=&
\ {\rm exp} \ -\frac{i}{\hbar}\; \left[ \int_0^t \ ds_1 \ \int_0^{s_1} \ ds_2
\ 2 \; q(s_1) \; Q(s_2) \; \eta(s_1 - s_2) \right] \nonumber \\
&\times&
\ {\rm exp} \ -\frac{1}{\hbar}\; \left[
\int_0^t \ ds_1 \int_0^{s_1} \ ds_2 \; q(s_1) \; q(s_2) \; \nu(s_1 - s_2)
\right],
\end{eqnarray}
where,
\begin{equation}
q(s) \ = \ x(s) \ - \ y(s)
\ \ \ \ \ \ \ \ \ \
Q(s) \ = \ \frac{1}{2}\; \left[ x(s) \ + \ y(s) \right]\ ,
\end{equation}
and the kernels in $(5)$ are given in terms of the spectral density as,
\begin{equation}
\nu(s) \ = \ \int_0^\infty \ d\omega \ J(\omega)
\ \cos \; \omega s
\end{equation}
and
\begin{equation}
\eta(s) \ = \ - \; \int_0^\infty \ d\omega
\ J(\omega) \ \tanh \; \frac{\hbar \omega}{2 k_B T}\; \sin \; \omega s.
\end{equation}
For this particular model the spectral density is written as,
\[
J(\omega) \ \equiv \ \sum_n \; \frac{J_n^2}{\hbar}\; \delta(\omega - \omega_n).
\]

Since the double path integral in $(4)$ is quadratic it can be evaluated
completely in terms of the initial and final coordinates given by the
transformation $(6)$, the result is :
\begin{eqnarray}
J(t,0) \ = \ Z(t) \ {\rm exp} \ \frac{i}{\hbar}
\biggl[ \ ( b_1 \; Q_0 &+& b_2 \; Q )\; q
\ - \ ( b_3 \; Q_0 + b_4 \; Q )\; q_0 \ \biggl] \nonumber \\
&\times& \ {\rm exp} \; - \frac{1}{\hbar}
\biggl[ a_{11} \; q^2 \ + \ a_{12}\; q_0 \; q \ + \ a_{22} \; q_0^2 \biggl]
\end{eqnarray}
where the coefficients $b_n$ ($n=1,...,4$) and $a_{nm}$ ($n,m=1,2$)
are given by
the solution of the equations :
\[
\frac{d^2 \kappa_n(s)}{d s^2} \
+ \ 2 \int^s_0{ds_1} \ \eta(s - s_1) \; \kappa_n(s_1) \ + \
\Omega_0^2\; \kappa_n(s) \ = \ 0
\]
with the boundary conditions,
\[
\kappa_1(s=0) \ = \ \kappa_2(s=t) \ = 1 \ \ \ \ \ \ \
\ \kappa_1(s=t) \ = \ \kappa_2(s=0) \ = 0.
\]
Then the unknown functions in $(9)$ are defined as,
\begin{eqnarray}
b_1(t) \ = \ M \; \frac{d\kappa_1}{ds}\, \biggl|_{s=t} \ \ \ \ \ \ \ \ \
b_2(t) \ = \ M \; \frac{d\kappa_2}{ds}\, \biggl|_{s=t}
\nonumber \\
b_3(t) \ = \ M \; \frac{d\kappa_1}{ds}\, \biggl|_{s=0} \ \ \ \ \ \ \ \ \
b_4(t) \ = \ M \; \frac{d\kappa_2}{ds}\, \biggl|_{s=0} \nonumber
\end{eqnarray}
and,
\[
a_{nm}(t) \ = \ \frac{1}{1 + \delta_{nm}} \ \int_0^t\ {ds_1}\ \int_0^{t}
\ {ds_2}
\ \kappa_n(s_1) \; \nu(s_1 - s_2) \; \kappa_m(s_2).
\]
Also the normalization factor in $(9)$ is given by,
\[
Z(t) \ = \ \frac{b_4}{2 \pi \hbar}.
\]

Using the above equations we can derive a set of identities,
\begin{equation}
\frac{\dot{b}_3}{b_4\; b_1} \ = \ -\frac{1}{M} \ \ \ \ \ \ \ \ \
\dot{a}_{22} \ = \ -\frac{\dot{b}_3\; a_{12}}{b_1} \ \ \ \ \ \ \ \ \
\frac{\dot{b}_4}{b_2\; b_4} =  \ -\frac{1}{M}
\end{equation}
(the dot over the functions denotes a derivative
with respect to $t$), which we shall
use latter.

The master equation is obtained using the simplified method proposed
by Paz \cite{JuanPablo}. Taking the derivative of
$(3)$ with respect to $t$ it is straightforward to show that,
\begin{eqnarray}
\frac{\partial \rho_s(q,Q,t)}{\partial t} \ &=& \
\left[ \frac{\dot{Z}}{Z} \; + \; \frac{i}{\hbar}\; \dot{b}_2 \; q \; Q
\; - \; \frac{1}{\hbar}\; \dot{a}_{11} \; q^2 \right]
\ {\rho}_s(q,Q,t) \nonumber \\
&+& \ \frac{i}{\hbar}\; \dot{b}_1 \; q
\ \int_{- \infty}^{+ \infty}\ dq_0
\ \int_{- \infty}^{+ \infty}\ dQ_0 \ Q_0 \ J(t,0) \ {\rho}_s(q_0,Q_0,0)
\nonumber \\
&-& \ \left[ \frac{i}{\hbar}\; \dot{b}_4\; Q + \frac{1}{\hbar}
\; \dot{a}_{12}\; q \right]
\ \int_{- \infty}^{+ \infty}\ dq_0
\ \int_{- \infty}^{+ \infty}\ dQ_0 \ q_0 \ J(t,0) \ {\rho}_s(q_0,Q_0,0) \\
&-& \ \frac{i}{\hbar}\; \dot{b}_3
\ \int_{- \infty}^{+ \infty}\ dq_0
\ \int_{- \infty}^{+ \infty}\ dQ_0 \ q_0 \ Q_0 \ J(t,0) \ {\rho}_s(q_0,Q_0,0)
\nonumber \\
&-& \ \frac{1}{\hbar}\; \dot{a}_{22}
\ \int_{- \infty}^{+ \infty}\ dq_0
\ \int_{- \infty}^{+ \infty}\ dQ_0 \ q_0^2 \ J(t,0) \ {\rho}_s(q_0,Q_0,0)
\nonumber
\end{eqnarray}

Although $(11)$ seems quite complicated we can simplify it by taking
derivatives of $(9)$ with respect to the end points, $q$ and $Q$,
which lead to the following relations,
\begin{equation}
q_0 \ J(t,0) \ = \ \left[ \frac{b_2}{b_4} \; q
+ i\hbar\; \frac{1}{b_4}\; \frac{\partial}{\partial Q} \right] \ J(t,0)
\end{equation}

\[
Q_0 \ J(t,0) \ = \ \left[ -\frac{b_2}{b_1}\; Q
- i \; \frac{a_{12}\; b_2}{b_1\; b_4}\; q
+ \hbar\; \frac{a_{12}}{b_1\; b_4}
\; \frac{\partial}{\partial Q}
-i \; \frac{2\; a_{11}}{b_1}\; q
- i \hbar \; \frac{1}{b_1}\; \frac{\partial}{\partial q} \right] \ J(t,0).
\]

By substituting $(12)$ in $(11)$ and using $(10)$ and $(3)$ we find after
some straightforward algebra that,
\begin{eqnarray}
i\hbar \, \frac{d{\rho}_s(x,y,t)}{dt} \ &=& \
\left[ -\frac{\hbar^2}{2M} \left(\frac{\partial^2}{\partial x^2}
- \frac{\partial^2}{\partial y^2}\right)
\ + \ \frac{M \, \Omega^2_R}{2}
\, (x^2 - y^2) \right] {\rho}_s(x,y,t) \nonumber \\
&-& i\hbar \ \Gamma(t) \, (x - y)\left(\frac{\partial}{\partial x} -
\frac{\partial}{\partial y}\right) {\rho}_s(x,y,t) \nonumber \\
&-& i \ D_{pp}(t) \, (x - y)^2 \, {\rho}_s(x,y,t) \\
&+& \hbar \ \Bigl[ \, D_{xp}(t) + D_{px}(t) \, \Bigl] \, (x - y)
\left(\frac{\partial}{\partial x} + \frac{\partial}{\partial y}\right)
{\rho}_s(x,y,t) \nonumber
\end{eqnarray}
is the master equation for the reduced density matrix of a particle
coupled to a two-level reservoir. Observe that the term which represents
the diffusion in real space, $D_{xx}$, is absent in this model since
the kernels in $(7)$ and $(8)$ depend only on the time interval
$s_1-s_2$ (see Eq. $(5)$). This is a general feature of stationary kernels
\cite{hu}.

The coefficients that appear in $(13)$
are given by,
\[
\Omega^2_R \ = \frac{\dot{b}_1\; b_2}{b_1} \; - \; \dot{b}_2
\]
is the renormalized frequency for the harmonic potential,
\begin{equation}
\Gamma(t) \ = \ b_2 \; + \; \frac{\dot{b}_1}{2\; b_1}
\end{equation}
is the dissipation coefficient,
\begin{equation}
D_{pp}(t) \ = \ \dot{a}_{11} \ - \ 4\; a_{11}\; b_2
\ + \ \dot{a}_{12}\; \frac{b_2}{b_4}
\ - \ \frac{\dot{b}_1}{b_1}\; \left[
2\; a_{11} \ + \ a_{12}\; \frac{b_2}{b_4} \right]
\end{equation}
is the diffusion coefficient in momentum space (it gives the fluctuation
in $\hat{x}^2$) and
\begin{equation}
D_{xp}(t) \ = \ D_{px}(t) \ = \ \frac{1}{4}
\ \left[ \frac{\dot{a}_{12}}{b_4} \ -
\ \frac{\dot{b}_1\; a_{12}}{b_1\; b_4} \ - \ 4\; a_{11} \right]
\end{equation}
is the diffusion coefficient which mix the real space and the
momentum space (it gives the fluctuation in $\hat{x}\hat{p}  +
\hat{p}\hat{x}$).

The first observation is that the kernels in $(7)$ and $(8)$ do not
have the same form as for the case of the oscillator bath \cite{caldeira}
unless we choose a {\it temperature dependent} spectral density of the form,
\begin{equation}
J_1(\omega) \ = \ \eta_1 \; \omega \ {\rm \coth}
\; \frac{\hbar \omega}{2 k_B T}
\ \theta(\omega_c - \omega)
\end{equation}
as it was first realized in Ref.\cite{twolevel} (it reduces to the spectral
density $(1)$ when $T \rightarrow 0$). With this choice the bath of oscillators
is indistinguishable from the bath of two-level systems. For example, in the
{\it high temperature} limit ($k_B T \gg \hbar \omega_c$), using (17)
we find,
\[
\nu(s) \ \approx \ \frac{2 \eta_1 k_B T}{\hbar}\; \delta(s)
\ \ \ \ \ \ \ \ \ \
\eta(s) \ = \ \eta_1  \frac{d \delta(s)}{d s}.
\]
In this case the diffusion and the dissipation are markovian while at
low temperatures the diffusion is clearly non-markovian. Moreover,
the coefficients that appear in $(14)-(16)$ are given by,
\begin{equation}
\Gamma(t) \ = \  \frac{\eta_1}{m}
\ \ \ \ \ \ \ \ \ \
D_{xp}(t)  \ = \ D_{px}(t) \ = 0
\ \ \ \ \ \ \ \ \ \
D_{pp}(t) \ = \ \frac{2 \eta_1 k_B T}{\hbar},
\end{equation}
as expected.

However, if we insist in using the {\it ohmic} spectral density, Eq.$(1)$,
the kernels in $(12)$ and $(13)$
assume, at {\it high temperatures} ($k_B T \gg \hbar \omega_c$), a non-local
form, namely,
\[
\nu(s) \ = \ - \; \eta_0 \; \frac{1}{s^2}
\ \ \ \ \ \ \ \ \ \
\eta(s) \ = \ \frac{\eta_0 \hbar}{k_B T} \; \frac{1}{s^3},
\]
and the dynamics is non-markovian. It is
interesting to notice that in the {\it low temperature}
limit ($k_B T \ll \hbar \omega_c$) the use of this ohmic spectral density
gives the following result for the kernels :
\[
\nu(s) \ = \ - \; \eta_0 \; \frac{1}{s^2}
\ \ \ \ \ \ \ \ \ \
\eta(s) \ = \ \eta_0 \; \frac{d \delta(s)}{d s},
\]
and the only time independent coefficient in the master equation is
the dissipation one : $\Gamma(t) = \frac{\eta_0}{m}$. This shows that
the two-level
system and
the oscillator bath cannot be naively mapped into each other with this
spectral density.

Another interesting possibility is a {\it flat} spectral density of the form,
\begin{equation}
J_2(\omega) \ = \ \eta_2 \; \theta(\omega_c-\omega),
\end{equation}
which is very common in applications in quantum optics \cite{louisell,pra}
(it should be noticed that $\eta_2$ is dimensionally different from
$\eta_0$, $\eta_1$).
In this case the diffusion is purely markovian (without memory) at any
temperature while the dissipation coefficient presents a transition from
markovian to non-markovian depending on the temperature scale, that is,
substituting $(19)$ in $(7)$ and $(8)$ and
taking the {\it high temperature}
limit, we find,
\[
\nu(s) \ = \ \eta_2 \; \delta(s)
\ \ \ \ \ \ \ \ \ \
\eta(s) \ \approx \ \frac{\eta_2 \hbar}{2 k_B T} \; \frac{d \delta(s)}{d s}.
\]
and the coefficients in the master equation are given by,
\[
\Gamma(t) \ = \ \frac{\eta_2 \hbar}{2 m k_B T}
\ \ \ \ \ \ \ \ \ \
D_{xp}(t) \ = \ D_{px}(t) \ = 0
\ \ \ \ \ \ \ \ \ \
D_{pp}(t) \ = \ \eta_2.
\]
Observe that in this case the dissipation coefficient has temperature
dependence (it decreases with the temperature) and the diffusion
coefficient is temperature independent. This result must be compared
with $(18)$ where the dissipation does not depend on the temperature
but the diffusion increases with the temperature.

In conclusion, in this paper we obtained the most general master
equation for a harmonic oscillator coupled to a reservoir consisting on
two-level systems. We show that the kernels which generate the diffusion and
dissipation coefficients in the master equation have a different form
from the standard cases of heat baths composed of harmonic oscillators.
We confirm the results of Caldeira {\it et al.} \cite{twolevel}
that the results obtained with an oscillator bath can be recovered
from a two-level system bath by a temperature dependent spectral density
(which reduces to the ohmic one when $T \rightarrow 0$).
We showed that the usual ohmic spectral
density produces a different kind of behavior from the usual oscillator model
\cite{caldeira}. We also studied the case of a flat spectral density (usually
used in quantum optics) which produces a different kind of
diffusion and dissipation for the quantum dynamics of the particle.

\acknowledgments
We would like to thank Amir O. Caldeira, Carlos O.Escobar and Andrew Matacz
for many helpful comments and suggestions. P.C.M. wishes to thank
Funda\c c\~ao de Amparo \`a Pesquisa no Estado de S\~ao Paulo
for financial support. A.H.C.N. acknowledges CNPq (Brazil) for
financial support. This research was supported in part by the
National Science Foundation under Grants No. PHY89-04035 and
DMR91-22385.

\end{document}